\documentclass[%
 reprint,
 amsmath,amssymb,
 aps,
]{revtex4-1}

\usepackage{graphicx}
\usepackage{dcolumn}
\usepackage{bm}

\usepackage[utf8]{inputenc}
\usepackage[T1]{fontenc}    
\usepackage{hyperref}       
\usepackage{url}            
\usepackage{amsfonts}       
\usepackage{nicefrac}       
\usepackage{microtype}      

\usepackage{lipsum}

\usepackage{mathtools}
\usepackage{amssymb,amsfonts}
\DeclarePairedDelimiter\abs{\lvert}{\rvert}

\usepackage{subfigure}
\usepackage{tikz}
\usetikzlibrary{decorations.pathmorphing, patterns,shapes}
\usepackage{wrapfig}

\renewcommand{\epsilon}{\varepsilon}
\usepackage{xcolor}

\begin{document}
\preprint{APS/123-QED}
\title{Nonlinear states and dynamics in a synthetic frequency dimension }
\author{Aleksandr K. Tusnin}
\email{aleksandr.tusnin@epfl.ch}
\author {Alexey M. Tikan}
\email{alexey.tikan@epfl.ch}
\author{Tobias J. Kippenberg}
\email{tobias.kippenberg@epfl.ch}
 \affiliation{Institute of Physics, Swiss Federal Institute of Technology Lausanne (EPFL), Lausanne, Switzerland}
\begin{abstract}

 Recent advances in the study of synthetic dimensions revealed a possibility to employ the frequency space as an additional degree of freedom which allows for investigating and exploiting higher-dimensional phenomena in a priori low-dimensional systems.
 However, the influence of nonlinear effects on the synthetic frequency dimensions was studied only under significant restrictions. In the present paper, we develop a generalized mean-field model for the optical field envelope inside a single driven-dissipative resonator with quadratic and cubic nonlinearities, whose frequencies are coupled via an electro-optical  resonant temporal modulation. The leading order equation takes the form of driven Gross-Pitaevskii equation with a cosine potential. We numerically investigate the nonlinear dynamics in such microring resonator with a synthetic frequency dimension in the regime where parametric frequency conversion occurs. We observe that the modulation brings additional control to the system, enabling one to readily create and manipulate bright and dark dissipative solitons inside the cavity. In the case of anomalous dispersion, we find that the presence of electro-optical mode coupling confines and stabilizes the chaotic modulation instability region. This leads to the appearance of a novel type of stable coherent structures which emerge in the synthetic space with restored translational symmetry, in a region of parameters where conventionally only chaotic modulation instability states exist. This structure appears in the center of the synthetic band and, therefore, is referred to as Band Soliton. Finally, we extend our results to the case of multiple modulation frequencies with controllable relative phases creating synthetic lattices with nontrivial geometry. We show that an asymmetric synthetic band leads to the coexistence of chaotic and coherent states of the electromagnetic field inside the cavity i.e. dynamics  that can be interpreted as chimera-like states. Recently developed $\chi^{(2)}$ microresonators can open the way to experimentally explore our findings.
\end{abstract}

\maketitle
\begin{figure*}
    \centering
    \includegraphics[width=\textwidth]{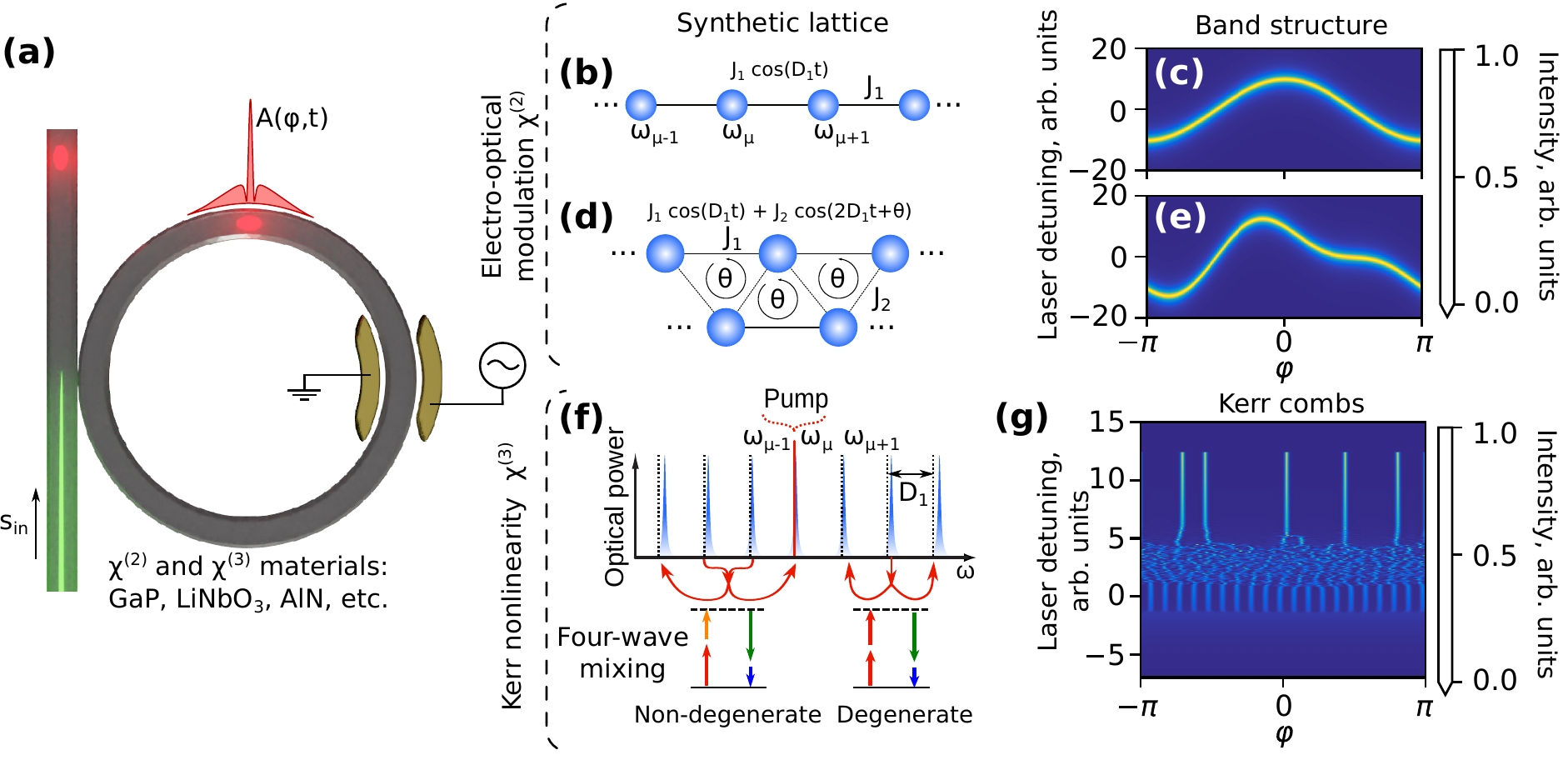}
    \caption{\textbf{Dynamically modulated optical cavity with $\chi^{(2)}$ and $\chi^{(3)}$ susceptibilities.} \textbf{(a)} Optical cavity with an integrated EO (electro-optical) modulator. The modulation frequency is integer number of FSRs (free spectral ranges): $\Omega = s D_1$, where  $s \in \mathbb {N}$. Due to the modulation, modes with frequencies $\omega_\mu$ and $\omega_{\mu+s}$ become coupled with coupling strength $J_s$ creating a synthetic lattice. \textbf{(b)} Schematics of the lattice with the nearest neighbor coupling ($s = 1$). \textbf{(c)} Corresponding cavity field response with $J_1 = 10 \kappa/2$, which represents the band structure. \textbf{(d,e)} the same as \textbf{(b,c)} but in the case of dual-tone modulation with relative phase $\theta = \pi/2$ and $J_2 = 0.45 J_1$.  \textbf{(f)} Displacement of the cavity resonance (in blue) from their exact equidistant positions (black dotted lines) due to the presence of dispersion. \textbf{(g)} Conventional nonlinear dynamics in a Kerr optical microresonator with anomalous group velocity dispersion.}
    \label{fig:pic1}
\end{figure*}
\section{Introduction}

The idea of unification of physical theories by using higher dimensional models beyond the usual space-time paradigm has arisen in the early years of development of quantum mechanics~\cite{klein1991quantum} and became an important precursor for modern unification theories~\cite{wesson2006five}. However, investigation of effects presented in higher dimensions  faces apparent challenges related to the number of dimensions provided by conventional physical systems. Boada and co-authors~\cite{Boada2012Quantum} have proposed to address these challenges by extending the well-established quantum simulator platform based on cold atoms with an additional \emph{synthetic dimension}. The essence of the proposed idea was to encode an additional dimension into another degree of freedom (atomic spin state in this case) in the way that effective Hamiltonian is analogues to a higher-dimensional one.

Since then, the concept of synthetic dimensions has been extended and used in various branches of physics~\cite{ozawa2019topological}. It acquired special significance in photonics, where it provides platform for exploring otherwise hardly accessible physical phenomena~\cite{aspuru2012photonic} and employment of synthetic dimensions allows for the dimensional extension employing only internal degrees of freedom of a system. This approach has been successfully applied to simulating particle random walk~\cite{Regensburger2011Photon}, effects of Bloch oscillations~\cite{wimmer2015observation}, unidirectional invisibility and unconventional reflection in parity-time symmetric systems~\cite{regensburger2012parity}, Anderson localization~\cite{vatnik2017anderson,pankov2019anderson}, etc. Recently, synthetic dimensions have been used in the studies of topological photonics~\cite{lu2014topological,khanikaev2017two,ozawa2019topological2}. Observation of a large variety of topological effects employing the synthetic frequency dimension has been proposed theoretically~\cite{Lin2016,Dutt2019} or realized experimentally~\cite{lustig2019photonic,Yang2020Mode,zhang2019photonic,Dutt2020}.

Synthetic dimensions in photonics can be realized using different physical mechanisms~\cite{yuan2018synthetic}. For example, coupled oscillating waveguides~\cite{lustig2019photonic}, pair of coupled unequal loops~\cite{schreiber2010photons} and phase modulation inside a ring cavity~\cite{ozawa2016synthetic} allows for encoding a synthetic dimension into spatial discrete models, arrival time of pulses, and resonator modes, respectively. We will refer to the latter case as \emph{synthetic frequency dimensions}. It can be created by inserting an electro-optical (EO) modulator into the ring resonator circumference~\cite{yuan2016photonic,yuan2016bloch,ozawa2016synthetic}. Modulating intracavity field at a frequency equal to an integer number of free spectral ranges (FSRs) (FIG.~\ref{fig:pic1}(a)), one can establish an effective photon flux between different optical modes supported by the resonator. In the case of the nearest-neighbor coupling (single FSR modulation) this system becomes similar to one-dimensional chain of identical atoms (see FIG.~\ref{fig:pic1}(b)). However, in contrast to Solid State Physics~\cite{ashcroft2016solid}, the modulated cavity modes play the role of a direct space, whereas time acts as a reciprocal one. Hence, exciting a cavity with an external laser which operates at a frequency $\omega_p$ close to the resonant $\omega_0$ and measuring the intracavity field response as a function of detuning $\omega_0 - \omega_p$, one can readily obtain a cosine-like band structure of the chain~\cite{ADutt2019Experimental} (see FIG.~\ref{fig:pic1}(c)). Furthermore, applying dual-tone modulation creates an effective two-dimensional frequency crystal (FIG.~\ref{fig:pic1}(d)) with controllable coupling strength (applied voltage) and phase flux (relative modulation phase) as introduced in the pioneering work by Dutt et al.~\cite{ADutt2019Experimental}. Due to non-zero phase flux, the corresponding band structure has non-reciprocal profile (FIG.~\ref{fig:pic1}(e)). 
   
Strikingly, the role of nonlinearity in photonic synthetic frequency dimension is hardly explored, however it is of particular importance for simulating locally interacting Hamiltonians~\cite{ozawa2017synthetic,Yuan2019Creating,Barbiero2019Bose} in complex many-body systems which are actively investigated in the context of  photonic quantum  simulators development~\cite{ozawa2019topological}. Yuan and co-authors have proposed a scheme that employs Kerr nonlinearity to achieve the local interaction between the simulated particles~\cite{Yuan2019Creating}. They have simulated a synthetic state governing by an effective Bose-Hubbard Hamiltonian and successfully explored the photon blockade effect. Even though this approach requires fulfillment of very restrictive conditions (such as zero dispersion and conserved total number of photons, which restrains this study to low-power regime), it is nonetheless very powerful since experimental platforms for realizing synthetic frequency dimensions often include materials with nonzero $\chi^{(3)}$ optical susceptibility. 

Remarkably, balance between cubic nonlinearity and dispersion (see FIG.~\ref{fig:pic1}(f)) allows generating different types of solitary waves, including solitons~\cite{Agrawal2000Nonlinear,KivsharOpticalSolitons}. In optical microcavities, an additional balance between parametric gain and cavity losses leads to the formation of \emph{dissipative} Kerr solitons (DKS) (FIG.~\ref{fig:pic1}(g)). Today, it is a very active field of research with wide raging applications~\cite{Lugiato2018,Lugiato1987Spatial,Kippenberg2018Dissipative}. Theoretically, dissipative solitons have been predicted in $\chi^{(2)}$ resonators as well~\cite{Buryak2002Optical,Leo2016Frequency}. Recent experimental observations with such photonic platforms as lithium niobate~\cite{Zhang2019,Ang2019,Gong2019}, aluminum nitride~\cite{Bruch2020Pockels}, and gallium phosphide~\cite{Wilson2020Integrated} along with new theoretical activities~\cite{Rowe2019Temproral,Podivilov2020Nonlinear,MasArab2020Modeling,Lobanov2020TwoColor}  in $\chi^{(2)}$ (and $\chi^{(2)}$-$\chi^{(3)}$) microresonators create a promising basis for the future development of this field and open new opportunities for investigation of nonlinear topological photonics~\cite{Yang2020Mode,Smirnova2019}.

The present paper investigates for the first time the nonlinear dynamics in a dispersive cavity with $\chi^{(2)}$ and $\chi^{(3)}$ optical susceptibilities where voltage-induced phase modulation creates a synthetic frequency dimension. Starting from the coupled-mode formalism, we derive mean-field Gross-Pitaevskii equation with a cosine potential which describes nonlinear dynamics of resonantly modulated intracavity field in microresonators and fiber loop cavities~\cite{Wabnitz1993Suppression}. We found that the modulation leads to predictable dissipative Kerr solitons (DKSs)~\cite{herr2014temporal,Kippenberg2018Dissipative} emergence and possibility of generation soliton crystals on-demand~\cite{cole2017soliton,he2019perfect,karpov2019dynamics}. We found that the modulation instability (MI) becomes  bounded by the curved bi-stability region. Surprisingly, with increasing of the coupling rate, new stable coherent structures emerge in the MI region, which we call \emph{Band Soliton}. These states appear to be dispersionless which makes them of particular interest in the context of synthetic frequency dimensions. Introducing a second tone to the intracavity phase modulation, we effectively create a nontrivial geometry which enables a nonreciprocal photon transfer~\cite{ADutt2019Experimental}. This leads to the coexistence of stable coherent and chaotic regions which we interpret as chimera-like states~\cite{Nielsen2019}. Our results highlight the rich Physics that can be accessed in synthetic dimensions with cubic nonlinearity.

\section{Theory}\label{sec:theory}
We consider an optical ring coupled to a bus waveguide with external coupling rate $\kappa_{\mathrm{ex}}$ (FIG.~\ref{fig:pic1}(a)). The cavity excited by a monochromatic laser with photon flux $s_{\mathrm{in}} = \sqrt{P/\hbar\omega_p}$ ($P$ is the input power) and frequency $\omega_p$, which is close to resonance frequency $\omega_0$. We suppose the modes being not equally spaced due to the dispersion, so the mode frequency ($\omega_\mu$) depends on the mode number ($\mu$) as $\omega_\mu = \omega_0 + D_1 \mu +\mu^2 D_2/2$, where $D_1/2\pi$ equals to FSR, and $D_2$ characterizes the group velocity dispersion (GVD) (FIG.~\ref{fig:pic1}(f)). A synthetic frequency dimension is created by an EO modulator at one part of the ring with modulation frequency $\Omega = s D_1$ with $s\in \mathbb{N}$~\cite{yuan2018synthetic}. Considering on-resonance coupling, only the modes with frequencies $\omega_\mu$  and $ \omega_{\mu\pm s} = \omega_\mu \pm s D_1 + \frac{D_2}{2}(s^2 \pm 2\mu s)$ interact. The dispersion shifts the resonance positions, leading to altering of nearest neighbors coupling efficiency. The linear equation of motion for the slowly varying mode amplitude $b_\mu$ can be written as
\begin{equation}\label{eq:CME_EOM}
    \frac{\partial b_\mu}{\partial t} = \frac{iJ_{s}}{2}\big(b_{\mu-s}e^{i\frac{D_2s}{2}(2\mu-s)t -i\theta} +b_{\mu+s}e^{-i\frac{D_2s}{2}(2\mu+s)t + i\theta} \big),
\end{equation}
where $J_s$ is the coupling rate with an arbitrary global phase $\theta$ correspond to modulation $J \cos{(s D_1 t + \theta)}$. By employing the Fourier transform of the field, one can deduce that this term may be rewritten as cosine potential for the electric field, so the governing equation for the electric field envelope in $\chi^{(3)}$ resonator under EO modulation takes the form of driven-damped Gross-Pitaevskii equation (GPE) (see Appendix~\ref{sec:ap1} for the derivation). In dimensionless form, it can be written as
\begin{widetext}
    \begin{equation}\label{eq:GPE_norm}
        \frac{\partial \Psi}{\partial \tau} = -(1 + i\zeta_0)\Psi + id_2\frac{\partial ^2 \Psi}{\partial \varphi^2 } + i\\|\Psi \\|^2\Psi + iJ\cos{(s\varphi + \theta)}\Psi + f,
    \end{equation}
\end{widetext}
where normalized variables $\tau = t/\tau_{\mathrm{ph}}$, $\tau_{\mathrm{ph}} = 2/\kappa$ is photon lifetime, $d_2 = D_2/\kappa$, $\zeta_0 = 2\delta\omega/\kappa$, $\delta\omega = \omega_0  -\omega_p$, $J = 2J_s/\kappa$,  $f = \sqrt{8\kappa_{ex}g_0/\kappa^3}s_{\mathrm{in}}$, $\Psi = \sqrt{2g_0/\kappa}A$, $\kappa = \kappa_{\mathrm{ex}}+\kappa_0$, $\kappa_0$ is intrinsic loss rate, $g_0$ is single-photon Kerr frequency shift, $A$ describes the optical field envelope and normalized such that  $\int_{0}^{2\pi} |A|^2 d\varphi/2\pi$ is the number of photons inside the cavity.
In the case of $J=0$ this equation corresponds to conventional LLE~\cite{Lugiato1987Spatial}. 

\begin{figure*}
    \centering
    \includegraphics[width=\textwidth]{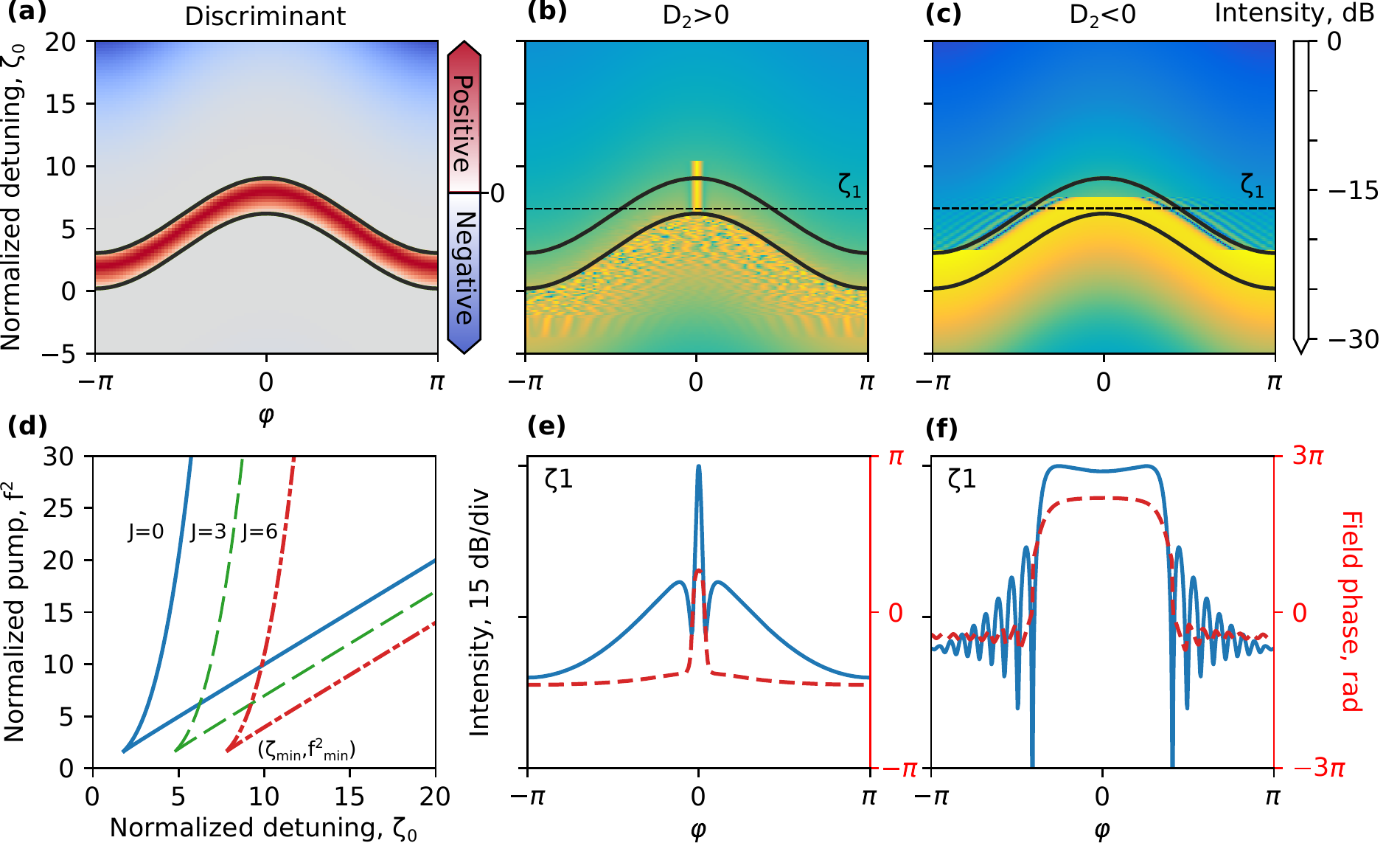}
    \caption{\textbf{Bi-stable branches and DKS (dissipative Kerr soliton)/platicon-existence range.} \textbf{(a)} Value of the discriminant $\Delta$ (Eq.~(\ref{eq:discr})) for coupling rate $J = 3$ and pump $f^2 = 6$ as a function of $\zeta_0$ and $\varphi$; \textbf{(b)} and \textbf{(c)} intracavity field in the case of anomalous and normal GVD (group velocity dispersion), respectively. Black solid lines represent value $\Delta = 0$ and indicate the bi-stable zone. A novel dynamics is observed in chaotic regime: MI (modulation instability) does not penetrate the bi-stable region. Soliton existence range is almost covered by the bi-stable region at $\varphi=0$. \textbf{(d)} Bi-stability range at $\varphi=0$ for coupling values $J = 0$ (solid), $J = 3$ (dashed), and $J = 6$ (dot-dashed). With increase of coupling $J$, the bi-stable zone shifts into the effectively red detuned region ($\zeta_0>0$) preserving its width. \textbf{(e)} Amplitude (solid blue) and phase (dashed red) profiles of DKS for detuning $\zeta_1 = 6.3$ (dashed lines on \textbf{(b,c)}).\textbf{(f)} The same in the case of normal dispersion.}
    \label{fig:pic2}
\end{figure*}

Let us begin with the analysis of stable solutions in the dispersionless limit ($d_2 = 0$). Introducing  $I = \abs{\Psi}^2$, one can readily derive the cubic equation
\begin{equation}\label{eq:NL_NoDisp}
    \Big(1 + \big(I + J \cos(s \varphi) - \xi_0\big)^2 \Big)I = f^2.
\end{equation}
The roots of this equation can be analyzed through its discriminant $\Delta$ (see Appendix~\ref{sec:ap2}). Depending on the sign of $\Delta$, there are three scenarios for solutions of Eq.~(\ref{eq:NL_NoDisp}): if $\Delta<0$ there is one real root and two complex conjugated roots, if $\Delta = 0$ roots are real and at least two of them are equal, if $\Delta > 0$ roots are real distinct numbers. Thus, negative (positive) discriminant corresponds to mono-stable (bi-stable) solutions, and in order to determine the bi-stability zone one needs to find $f^2$ and $\zeta_0$ such that the discriminant equals to zero. Since Eq.~(\ref{eq:NL_NoDisp}) explicitly depends on $\varphi$, the discriminant becomes $\varphi$ dependent, therefore \emph{different spatial parts} of the cavity are found at \emph{different parts of the stability diagram at the same value of laser detuning} (see FIG.~\ref{fig:pic2}(a)). 

We note that a similar effect can be achieved imposing external resonant modulation on the pump laser~\cite{Nielsen2019,Anderson2019}. External modulation has been employed for DKSs locking and manipulations creating an effective potential that traps DKSs~\cite{Obrzud2017}, it also helps to trigger platicon generation~\cite{Lobanov2015}. However, the $\varphi$ dependence lies in the right hand side of Eq.~(\ref{eq:NL_NoDisp}). Therefore, it is expected that the internal phase modulation will provide an additional degree of freedom for controlling emergent coherent structures as well.  

The threshold value $f^2$ which corresponds to the triple real root of Eq.~(\ref{eq:NL_NoDisp}) can be obtained analytically, and it equals to $f^2_{min} = 8\sqrt{3}/9$, which coincides with the critical value for the resonance tilt for LLE~\cite{Godey2014}. Remarkably, this result does not depend on $\varphi$, despite the $\varphi$ dependence of Eq.~(\ref{eq:NL_NoDisp}).

\section{Numerical simulations}\label{sec:num_sim}
\subsection{Dynamics of dissipative solitons and platicons}
For further analysis we consider the case $s = 1$. In FIG.~\ref{fig:pic2}(a) we show the values of the discriminant $\Delta$ as a function of $\varphi$ and $\zeta_0$ for pump rate $f^2 = 6$ and coupling $J = 3$. As one can see, the presence of the potential leads to bending of the bi-stable zone in a way, that for a given detuning the system can be simultaneously on the mono-stable  and  bi-stable branches. With increasing of coupling strength $J$, the bistability zone bends further and goes deeper into the effectively red detuned region $(\zeta_0 >0)$ (see FIG.~\ref{fig:pic2}(a)). 

\begin{figure*}
    \centering
    \includegraphics[width=\textwidth]{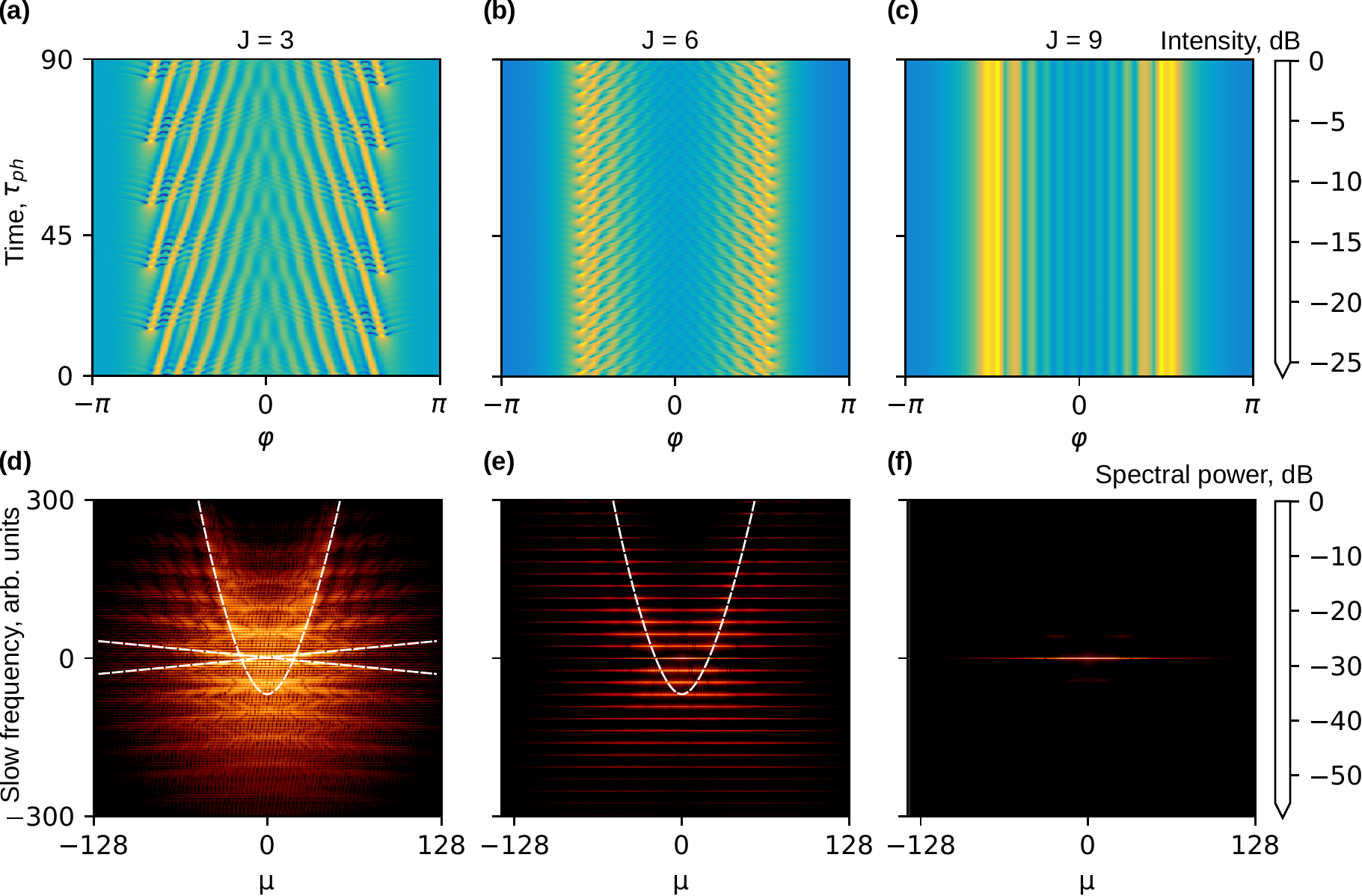}
    \caption{\textbf{Dynamics of the confined MI (modulation instability) region in the presence of potential}. \textbf{(a-c)} Spatio-temporal diagrams of the intracavity field for $f^2 = 6,\, \zeta_0 = 1.28$.  Coupling $J = 3$ corresponds to \textbf{(a)},  $J=6$ \textbf{(b)}, and $J = 9$ \textbf{(c)}. \textbf{(d-f)} Corresponding NDR (nonlinear dispersion relation) which represents effective nonlinear dispersion relation of the system. One can notice how the dispersion relation transforms with increase of $J$. For $J=3$ the system consists of constantly appearing and colliding dispersionless structures (lines with opposite slopes in \textbf{(d)}) which radiate dispersive waves (parabola in \textbf{(d)}), for $J=6$ \textbf{(b,e)} the field oscillates as a whole and forms a ladder in the NDR profile, which indicates periodic breathing in time. Further increase of $J$ \textbf{(c,f)} transforms the field into dispersionless stable dissipative structure. }
    \label{fig:pic3}
\end{figure*}

We continue the further analysis by performing numerical simulation of GPE~(\ref{eq:GPE_norm}), taking $d_2 = \pm 0.01$ and scanning the cavity from blue-  ($\zeta_0<0$) to red-detuned side. We employ numerical integration utilizing the split-step Fourier method~\cite{Agrawal2000Nonlinear}. The positive (negative) value of $d_2$ corresponds to anomalous (normal) dispersion regimes.  We analyze these cases separately. 
\paragraph{Anomalous dispersion.}
We observe that the presence of the potential in GPE~(\ref{eq:GPE_norm}) breaks the translational symmetry along  $\varphi$ coordinate and leads to confinement of the MI region~\cite{Nielsen2019}. We observe that chaotic patterns do not penetrate into bi-stable zone, and DKS appear at the center of the cavity (FIG.~\ref{fig:pic2}(b)). The latter might be qualitatively understood through the analysis of the steady-state dispersionless linear solution, which can be considered as a background for the dissipative nonlinear structures in the cavity. The intracavity field can be expressed as
\begin{equation}\label{eq:exact_sol_lin_nd}
    \Psi = \frac{f}{1+i(\zeta_0 - J \cos(\varphi))}.
\end{equation}
Depending on normalized detuning, the field intensity has one ($\zeta_0>J$, $\varphi_0 = 0$) or two ($\zeta_0<J$, $\varphi_{\pm} = \pm \arccos{\xi_0/J}$) maxima. When the modulated background  has only one peak, a single DKS can be formed on it. Numerical simulations show that the DKS appears on the peak of the modulated background in the bi-stable region (FIG.~\ref{fig:pic2}(b))~\cite{Jang2015}. The width of this region as a function of $f^2$ and $J$ can be calculated analytically (see Appendix~\ref{sec:ap2} for the details), and we present it on the FIG.~\ref{fig:pic2}(d) for coupling rates $J = 0,\,3,\,6$. Surprisingly, this zone simply shifts into the effectively red-detuned region linearly with $J$, and the critical detuning for $f^2_{\mathrm{min}}$ is 
\begin{equation}
\zeta_{\mathrm{min}} = \sqrt{3}+J. 
\end{equation}

In order to calculate the soliton existence range, we employ Lagrangian perturbative approach~\cite{grelu2015nonlinear,Wabnitz1993OpticalMem}. First of all, we introduce the change of variable $\Theta = 1/\sqrt{2d_2} \varphi$ to the equation~(\ref{eq:GPE_norm}). Thus, the equation for the Lagrangian density can be written as follows: 
\begin{align}\label{eq:Lagr_dens}
    \mathcal{L} =& \frac{i}{2}\big(\Psi^*\frac{\partial \Psi}{\partial \tau} - \Psi\frac{\partial \Psi^*}{\partial \tau} \big) - \frac{1}{2}\Big|\frac{\partial \Psi}{\partial \Theta}\Big|^2 +\nonumber\\
    &+\frac{1}{2}\abs{\Psi}^4 + (J\cos(\alpha \Theta) - \zeta_0)\abs{\Psi}^2,
\end{align}
 where $\alpha = 2d_2$. The dissipative function is introduced in the form:
\begin{equation}\label{eq:dis_func}
    \mathcal{R} = -i \Psi + if.
\end{equation}
The Lagrangian $L = \int \mathcal{L}d \Theta$ obeys:
\begin{equation}\label{eq:lagr_eq}
    \frac{\partial L}{\partial q_i}- \frac{d}{d \tau}\frac{\partial L}{\partial \dot{q}_i} = \int \Big(\mathcal{R}\frac{\partial \Psi^*}{\partial q_i} + \mathcal{R}^*\frac{\partial \Psi}{\partial q_i}  \Big)d\Theta.
\end{equation}
Using the ansatz of a stationary soliton $\Psi=B \text{sech}(B\Theta)e^{i\xi_0}$ and considering $q_1 = B$ and $q_2 = \xi_0$, we get (see Appendix~\ref{sec:ap3} for more details)
\begin{align}
    \frac{d B}{d \tau} &= -2B + \pi f \cos\xi_0, \label{eq:B_t} \\
    \frac{d \xi_0}{d\tau} &= \frac{1}{2}B^2 - \zeta_0 + J \delta^2 \frac{\cosh \delta}{\sinh \delta}\frac{1}{\sinh\delta},\label{eq:xi_t}
\end{align}
where we define $\delta = \alpha \pi/2B$. Considering $\alpha \ll 1$ and using Tailor expansion we obtain the stable solution
\begin{align}
    B^2 &= 2(\zeta_0 - J),\\
    \cos \xi_0 &= \frac{2 B}{\pi f}.
\end{align}
From the latter we obtain the analytical expression of the maximum detuning for stable soliton in the presence of nearest-neighbor coupling 

\begin{equation}
\zeta_{\mathrm{max}} = \pi^2f^2/8+J.
\end{equation}
This result generalizes the known expression for the soliton existence range to single-tone EO modulated cavity. Similarly to bi-stable zone, the maximum detuning $\zeta_{\mathrm{max}}$ simply shifts by $J$.

Increasing the modulation frequency (i.e. increasing of $s$ in Eq.~(\ref{eq:GPE_norm})) leads to period multiplication of the modulated background and allows for creating soliton crystals~\cite{Cole2017,karpov2019dynamics} with $s$ equally spaced DKSs. Alternatively, applying several modulation signals and having control of the modulation phase, one can control positions and the number of DKS in the cavity, which enables controlled soliton tweezing~\cite{Jang2015}, and as shown later, leads to a new dynamics.

\paragraph{Normal dispersion.}
In the context of the conventional LLE with normal GVD ($d_2 < 0$), the dark solitons (also called platicons) are hard to excite by simple laser tuning (soft excitation) for relatively small detunings and pump rates~\cite{Godey2014}. In order to create them, one needs to use additional methods, such as pumped modulation~\cite{Lobanov2015}, or pumping in the avoided mode crossings~\cite{Xue2015,Kim2019}. In this context, EO modulation provides with an effective flux of photons from the pumped resonance to sidebands, making platicons accessible without additional perturbations. In FIG.~\ref{fig:pic2}(c) one can see platicon generation in the resonantly modulated cavity. In contrast to the DKS, the platicons appear only when one part of the cavity passes the whole bi-stable region; however, FIG.~\ref{fig:pic2}(d) can still indicate approximate platicon existence range.

\begin{figure*}
    \centering
    \includegraphics[width=\textwidth]{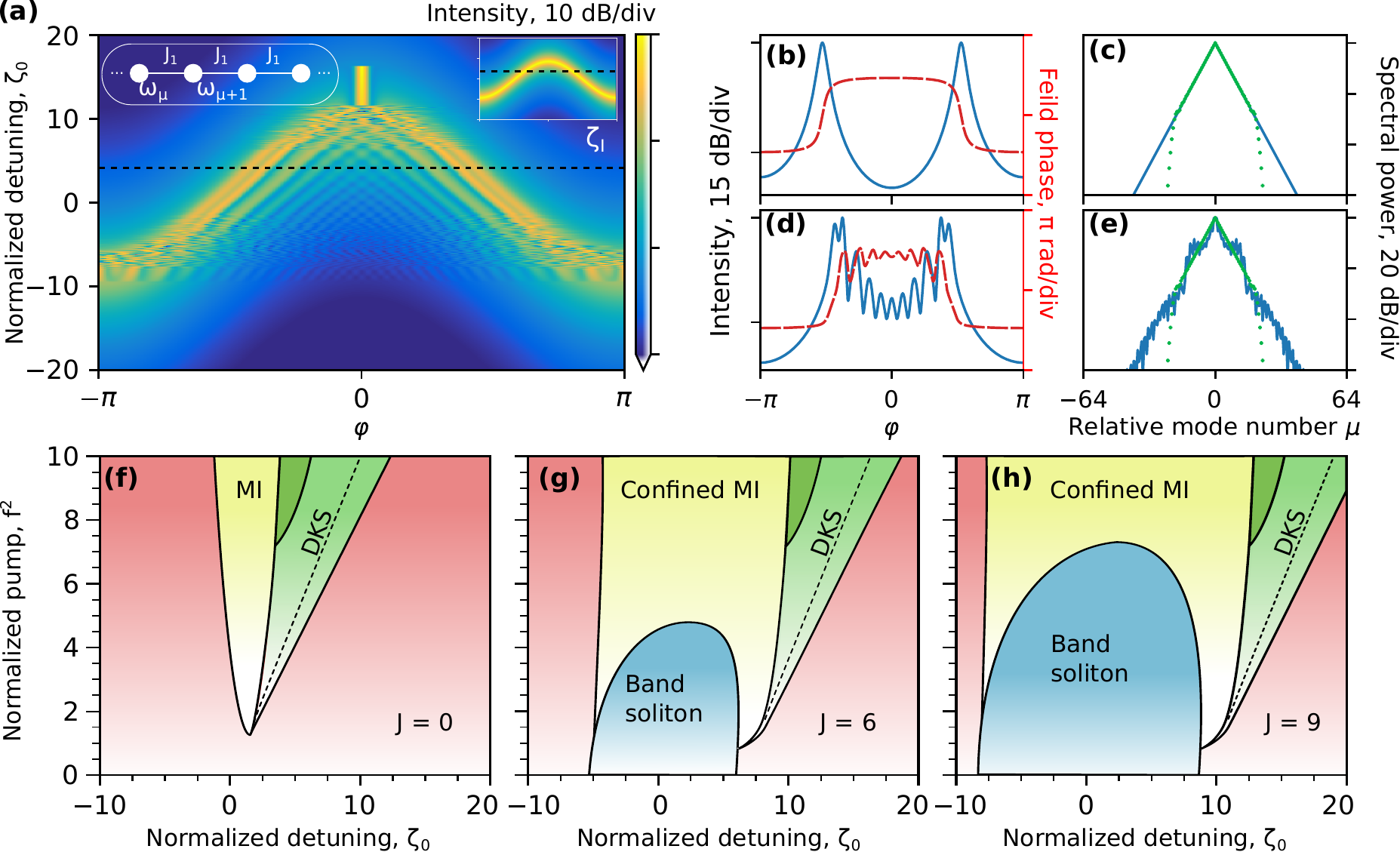}
    \caption{\textbf{New nonlinear states termed Band Solitons for single-tone modulation.} \textbf{(a)} Intracavity field for  potential $J \cos \varphi$, coupling strength $J = 9$, and pump rate $f^2 = 6$. Band solitons emerge in the detuning range $\zeta_0 \in [-3.5,7]$. Insets show the linear dispersionless case: schematics of the lattice for coupling $J \cos \varphi$ in linear case without dispersion (upper left) and  corresponding cavity response (upper right). Horizontal line corresponds to detuning $\zeta_{\text{I}} = 4$, for which we examine field and spectrum profiles (\textbf{b-e}). \textbf{(b)} Linear field intensity (solid blue line) and phase (dashed red line).  \textbf{(c)} Linear field spectrum without dispersion (solid blue line) and with dispersion $d_2=0.01$ (dotted green line). \textbf{(d)} Nonlinear field intensity (solid blue line) and phase (dashed red line).  \textbf{(e)} Nonlinear field spectrum (solid blue line) and corresponding linear spectrum with dispersion (dotted green line). \textbf{(f-h)} Phase diagrams for coupling strengths $J = 0$, $J = 6$ and $J = 9$. The red zone corresponds to continuous wave; the yellow zone indicates the confined MI (modulation instability) state; the green zone corresponds to the soliton existence range, which is predicted analytically. The dashed line indicates the end of the bi-stable region. The dark green region depicts DKS breathers. The blue zone indicates existence range of \emph{band solitons}, a new type of dissipative coherent structure that appears in a conventional ($J=0$) chaotic MI region. }
    \label{fig:pic4}
\end{figure*}

\subsection{Confined MI region}
Let us restrict our consideration for the case of anomalous GVD ($d_2>0$). In the conventional LLE formalism, in order to generate DKS via the soft excitation, one needs to scan the resonance through the MI region. In this region, coherent structures randomly appear and collide with each other, and may give birth to rogue waves~\cite{Coulibaly2019}. However, due to the modulated background, the nonlinear structures appear and interact at different parts of the resonator differently. In order to investigate the role of coupling $J$, we explore spatio-temporal diagrams at a fixed pump rate, detuning and coupling rate and its nonlinear dispersion relation (NDR), which is essentially the Fourier transform of the spatio-temporal diagram along two axes: time ($t$) and space ($\varphi$). This two-dimensional Fourier transform gives information about effective NDR and reveals insights about linear dispersive and nonlinear waves in the system~\cite{Leisman2019}. For instance, a single DKS will be presented as a line in this diagram with a slope, which indicates its group velocity; a breather is similar to DKS, but its profile consists of equally spaced lines, which indicate its breathing oscillation in time; the linear waves, which obey the linear dispersion law lie along the parabola. One may see all these features in the FIG.~\ref{fig:pic3}(a--f). First of all, we chose simulation parameters as in FIG.~\ref{fig:pic2}(b), but with fixed detuning $\zeta_0 = 1.3$. On the spatio-temporal diagram FIG.~\ref{fig:pic3}(a) one can see how nonlinear structures periodically arise and oscillate in the viscinity of background maxima $\varphi_\pm$, propagate towards the maximum of the background phase at $\varphi \approx 0$ (red dashed line in FIG.~\ref{fig:pic4}(b)) and annihilate. There are several distinct structures on the corresponding NDR (FIG.~\ref{fig:pic3}(d)): the periodic lines along the slow frequency axis with opposite slopes correspond to the colliding structures which locally have conventional DKS (dissipative Kerr soliton) profile; the parabola corresponds to dispersive waves which are emitted by the breathing DKS on the background. With increasing of the detuning these structures come closer, get smaller group velocity, and interact more chaotically while the field in the vicinity $\varphi = \pm \pi$ rests unperturbed. Thus, we observe that for relatively small coupling rates the potential leads to confinement of the MI (modulation instability) state. 

However, with increasing coupling strength ($J = 6,\,9$), we observe how this constantly interacting solitons are transformed into a new stable dispersionless structure (FIG.~\ref{fig:pic3}(b,c,e,f)). For coupling rate $J = 6$, we observe that the field starts to periodically oscillate in time. The corresponding NDR consists of a ladder of lines, which signifies the appearance of a new dispersionless breathing structure. Further increasing of the coupling ($J=9$) stabilizes this structure, it becomes coherent and \emph{dispersionless}. In the following sections we further explore this novel state.

\subsection{Band Soliton}
With increase of coupling strength $J$, we observe that the \emph{MI region is getting stabilized at a certain detuning range}, and \emph{new stable (i.e. coherent) nonlinear structures emerge}. In a linear dispersionless case with the nearest-neighbor coupling ($s = 1$), the intracavity field response for different detunings represents a band structure of a one-dimensional synthetic crystal. However, the presence of FWM introduces global nonlinear coupling between the modes, which efficiency is given by the chromatic dispersion. The latter signifies that the eigenfunction basis (see Ref.~\cite{ADutt2019Experimental}) is modified, and the intracavity field response can no longer be considered as a band structure. 

The FIG.~\ref{fig:pic4}(a) demonstrates this difference: the deterministic dispersionless response (see upper right inset) transforms into a complex structure, which contains localized chaotic and stable states. However, the notion of band structure remains important even in the nonlinear regime~\cite{Lumer2016,Solnyshkov2017}. We observe that there is a threshold value of $J$ for a given pump rate $f$ when the novel coherent structures appear. Comparing nonlinear response (FIG.~\ref{fig:pic4}(a)) with dispersionless linear one (FIG.~\ref{fig:pic4}(a), inset), we notice that these structures emerge in the center of the band structure, thus we call them \emph{band solitons}. In analogy to solid state physics, we can introduce the notion of synthetic Bloch waves (BW)~\cite{ADutt2019Experimental}, existing in the frequency space. Their group velocity reaches its maximum in the part of the band structure with the highest slope steepness. The latter signifies that the stable nonlinear states appear due to the interplay between FWM and linear BW. When the coupling strength is smaller than the threshold value, linear waves do not have sufficient velocity to redistribute perturbations induced by FWM. This regime corresponds to the confined MI. However, when the coupling strength exceeds the threshold value, the group velocity of the BW in the center of the band increases as well, and the BW can propagate faster along the frequency space and redistribute perturbations induced by FWM, leading to locking between the modes and the emergence of new coherent states. This reasoning can also be applied to the explanation of the conventional DKS states existence. As we have shown in previous sections, DKS appears exactly at the top of the band structure, where the group velocity of the BW equals to zero; hence the photon flux from the pump is provided only due to FWM, and the synthetic BWs do not affect this process. Due to this fact, this soliton corresponds to conventional soliton in optical $\chi^{(3)}$ microcavities.

Now we investigate the field's amplitude, phase and spectrum at $\zeta_{\text{I}} = 4$ (FIG.~\ref{fig:pic4}(d,e). In the linear dispersionless case, the solution can be found analytically (Eq.~(\ref{eq:exact_sol_lin_nd})), and the field incorporates two maxima (FIG.~\ref{fig:pic4}(b)). Corresponding spectral profile (FIG.~\ref{fig:pic4}(c), solid blue line) decays exponentially with mode number $\mu$, showing that  the coupling rests the same for all the modes. 
Dispersion shifts the modes, decreasing coupling efficiency for higher-order modes and leading to truncation of the spectrum and emergence of a conventional EO comb (green dots on the FIG.~\ref{fig:pic4}(c,e))~\cite{Zhang2019}. However, FWM shifts the resonances, enhancing coupling between the modes by restoring translational symmetry in the frequency space (see FIG.~\ref{fig:pic4}(e), solid blue line). The spectrum of this new state incorporates a flat part near the pump (at $-10$ dB level) and decays slower than the EO comb, which signifies the restored coupling between the modes beyond the cut-off~\cite{Zhang2019,Ho1993Optical,Kourogi1993WideSpan}.

 In order to investigate stability of these states, we scan the cavity for different pump rates and coupling strengths. In FIG.~\ref{fig:pic4}(g,h) we present the phase diagram for single-tone modulation with coupling strengths $J = 6$ and $J=9$ respectively and compare it with the conventional LLE model (FIG.~\ref{fig:pic4}(f)). The presence of the potential significantly changes the system dynamics, especially the MI region has new features. Band solitons emerge in a region around $\zeta_0 = 0$. With increasing $J$, their existence range increases along both axes. One can notice that this existence range is asymmetric, while in the linear case the band structure is symmetric (FIG.~\ref{fig:pic4}(a) upper right inset). However, FWM induces self-phase modulation, leading to the frequency shift towards the effectively red-detuned zone, and the whole band obtains an offset from $\zeta_0 = 0$. The band solitons transform to conventional EO combs at the low pump rates when FWM process becomes negligible. With increase of the pump rate, the band solitons start to breath, become unstable and transform to confined MI. Since the transition from the breathing state to the chaotic one is smooth, we joined these regions and labeled them as \emph{confined MI} in FIG.~\ref{fig:pic4}(g,h) (note, we do not indicate here a narrow region of stable MI, which always manifest itself at negative values of detuning). This region appears to be wider than MI region in the conventional LLE model (FIG.~\ref{fig:pic4}(f)).

\subsection{Chimera-like states}

\begin{figure*}
    \centering
    \includegraphics[width=\textwidth]{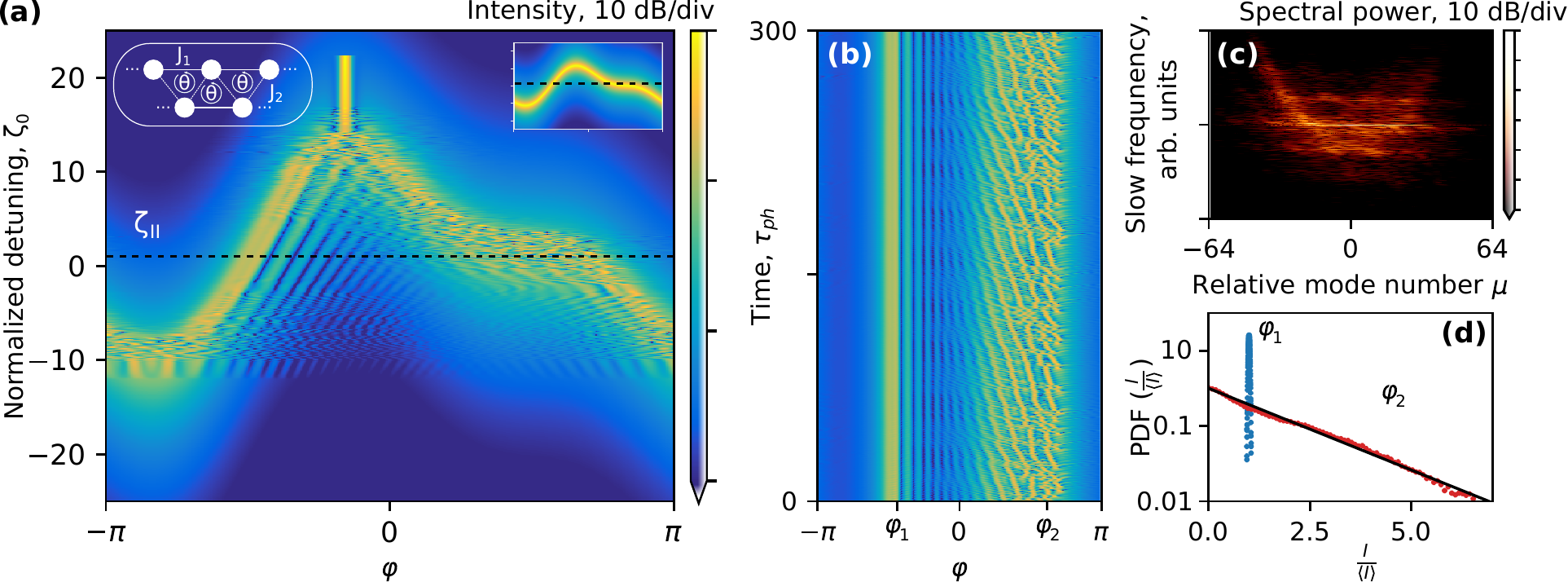}
    \caption{\textbf{Appearance of chimera-like states in the case of dual-tone modulation.} \textbf{(a)} Intracavity field for potential $J \big(\cos \varphi + 0.45 \cos (2\varphi + \theta)\big)$, coupling strength $J = 9$ with relative phase $\theta = \pi/2$ and $f^2 = 9$.  Insets: corresponding schematics of lattice (upper left) in linear case without dispersion and the cavity response (upper right). Horizontal line corresponds to detuning $\zeta_{\text{II}} = 1.9$, for which we examine spatio-temproal diagram \textbf{(b)}, NDR (nonlinear dispersion relation) \textbf{(c)}. \textbf{(d)} Single point PDF (probability density function) of the normalized intensity for two intracavity coordinates $\varphi_1 = -1.4$ (blue) and $ \varphi_2=1.8$ (red). Black solid line corresponds to exponential PDF $\exp{-\frac{I}{\langle I \rangle}}$.
    }
    \label{fig:pic5}
\end{figure*}

Using two modulation frequencies and controlling the relative phase between them, one introduces a two-dimensional synthetic lattice~\cite{ADutt2019Experimental} in the frequency space (FIG.~\ref{fig:pic5}(a) upper left inset). The phase flux between the nodes can be controlled in this arrangement by the relative modulation phase. In particular, one can obtain asymmetric band structure introducing nonreciprocal frequency conversion (FIG.~\ref{fig:pic5}(a) upper right inset)~\cite{ADutt2019Experimental}. We investigate nonlinear dynamics for a dual-tone modulation corresponding to the effective potential $J \big( \cos \varphi + 0.45 \cos (2\varphi + \theta)\big)$ with coupling $J = 9$ and the relative phase $\theta = \pi/2$. Nonreciprocal photon flow introduces a significant asymmetry in the corresponding spectral profile (FIG.~\ref{fig:pic5}(c))~\cite{Tzuang2014}. However, in contrast to the single tone modulation, it is possible to find a region where one side of the band structure is almost flat while another one has a maximum of its slope ($\zeta_0 \approx 2$ in FIG.~\ref{fig:pic5}(a)). Therefore, for certain coupling rates fully chaotic dynamics manifests itself in a part of the cavity where the synthetic band structure slope (and hence the photon flow due to the linear BWs) is small, while another side can support a novel coherent band soliton existence. A similar intriguing feature was recently observed in systems with local coupling~\cite{Clerc2017,Nielsen2019}. Following these works, we refer to the observed phenomenon as \emph{chimera-like state}.

In order to investigate the chimera-like state, we extract the complex field envelope at the detuning value $\zeta_{II} = 1.9$ (black dashed line in FIG.~\ref{fig:pic5}(a)) and numerically propagate fixing all the parameters. The dynamics of the field modulus is shown in FIG.~\ref{fig:pic5}(b). Nonreciprocal photon transfer breaks the underlying symmetry of the system which also follows from the NDR (see FIG.~\ref{fig:pic5}(c)). Computing the single point probability density function (PDF) of the intensity variation $I/\langle I \rangle$ ($\langle I\rangle$ is averaged intensity in time) in coherent ($\varphi_1 = -1.4$) and incoherent ($\varphi_2 = 1.8$) regions using $3\times 10^{5}$ samples, we show that the PDF at $\varphi_2$  approaches the exponential (i.e. Gaussian distribution for the real part of the field) which can be considered as a signature of a fully developed MI stage~\cite{Coulibaly2019} (also~\cite{agafontsev2015integrable,kraych2019statistical}), while  at $\varphi_1$ it is close to delta-like distribution. Such states have no counterparts in DKS-based on $\chi^{(3)}$ and single-tone driving.

\section{Conclusion}\label{sec:concl}
In summary, we proposed a theoretical model which describes nonlinear dynamics of a modulated optical cavity with $\chi^{(2)}$ and $\chi^{(3)}$ optical susceptibilities and second order GVD.  We have shown that in the linear dispersionless limit the model describes the physics of a ring with a synthetic frequency dimension.
Considering the dynamics of the full model, we found that despite the presence of GVD which breaks the translational symmetry there are coherent dispersionless structures for which the coupling remains resonant. There are two types of structures we have observed. First is found in the region of zero group velocity of the synthetic Bloch waves. They correspond to conventional DKS solutions of LLE but living on a modulated background. Applying different modulation signals, one can directly control the background modulation, hence control number and positions of DKS, making soliton crystals and soliton tweezing readily accessible. The second type of the structures is found at the maximum of the synthetic Bloch waves group velocity. Coherence of these novel structures, that we called Band Solitons, relies on the efficient photon transfer due to the linear mode coupling and Kerr nonlinearity which compensates the effect of dispersion. Therefore, such structures can be considered as \emph{nonlinear states in the synthetic frequency dimension}. We generalized this result by including far neighbor coupling (double-tone modulation) into the model. We found that due to the nonreciprocal photon transfer the symmetry of the system is broken which leads to the coexistence of stable coherent structures and chaos. We interpret these as the appearance of chimera-like states in the system. 

We would like to emphasize that the proposed model can be used for further investigation of the synthetic frequency dimension as well as for simulations of EO combs in $\chi^{(2)}$ resonators. It can be readily generalized for an arbitrary dispersion profile, which can incorporate either higher order dispersion $D_3,\, D_4$, or avoided mode crossings. Also, the potential of this model in the investigation of nonlinear effects in the synthetic frequency dimension in resonator lattices is of high interest. For example, by simulating a set of coupled GPE, one may explore the nonlinear dynamics of topological states, that can be created by changing the relative modulation phase of each ring.

As a physical platform for the model one can consider a high-Q optical microcavity with $\chi^{(2)}$ and $\chi^{(3)}$ optical susceptibilities. With recent success in fabrication process it has become possible to create optical cavities based on lithium niobate~\cite{Ang2019,Zhang2019} or aluminum nitride~\cite{Bruch2020Pockels}, as well as gallium phosphide~\cite{Wilson2020Integrated} photonic platforms. These materials are of particular interest
because they possess both quadratic and cubic susceptibilities, and it has been successfully used for generation of Kerr-based~\cite{Ang2019,Gong2019}, EO-combs~\cite{Zhang2019}, and Pockels soliton~\cite{Bruch2020Pockels}.
       
\section{Acknowledgments}
The authors thank J. Riemensberger, M. Karpov and M. Churaev for fruitful discussions. This publication was supported by contract D18AC00032 (DRINQS) from the Defense Advanced Research Projects Agency (DARPA), Defense Sciences Office (DSO). A.K. Tusnin acknowledges support from the European Union’s Horizon 2020 research and innovation programme under the Marie Skłodowska-Curie grant agreement No. 812818 (MICROCOMB).

\appendix
\section{Derivation of Gross-Pitaevskii equation }\label{sec:ap1}
Let us consider a ring resonator with a phase modulator distributed at one part of the ring. If modulator consists of $\chi^{(2)}$ active material, then it changes locally refractive index $n(\phi,t)$ and provides with linear coupling between different modes, which can be described by the equations of motion for the amplitudes $a_\mu$ as (see Supplementary Note 1 in Ref.~\cite{ADutt2019Experimental}),
\begin{equation}\label{eq:CME_EO}
    \frac{\partial a_\nu(t)}{\partial t} = -i \omega_\nu a_\nu(t) + i\sum_{\mu}J_{\mu-\nu}(t)a_\mu(t).
\end{equation}
Let us suppose that the coupling coefficient does not depend on $\mu$ and depends harmonically on time as $J_{\mu-\nu} = J_s \cos{(\Omega t + \theta)}$, where $\Omega$ is the modulation frequency and $\theta$ is the modulation phase. Under the transformation into rotating frame ($a_\nu = b_\nu e^{-i\omega_\nu t}$), the equation reads
\begin{equation}\label{eq:CME_EO_rot_frame}
    \dot{b}_\nu e^{-i\omega_\nu t} = \frac{i}{2}\sum_\mu J_{s} b_\mu e^{-i\omega_\mu t}\big(e^{i\Omega t + i\theta} + e^{-i\Omega t - i\theta} \big)
\end{equation}
($\dot{b}_{\mu}$ indicates time derivative).
Supposing that we pump the resonator near to frequency $\omega_0$ such that 
\begin{equation}
    \omega_\mu = \omega_0 + \mu D_1 + D_2\frac{\mu^2}{2},
\end{equation}
and modulation frequency $\Omega$ is chosen as $\Omega = s D_1$, where $s$ is an integer, the RHS of Eq.~(\ref{eq:CME_EO_rot_frame}) incorporates two exponentials
\begin{align}
    \omega_\mu+sD_1 - \omega_\nu &= D_1(\mu+s-\nu) + \frac{D_2}{2}(\mu^2 - \nu^2),\\
    \omega_\mu-sD_1 - \omega_\nu &= D_1(\mu-s-\nu) + \frac{D_2}{2}(\mu^2 - \nu^2).
\end{align}
Therefore, the resonant interaction appears between modes $\mu = \nu \pm s$, and the equation~\ref{eq:CME_EO_rot_frame} simplifies to
\begin{equation}\label{eq:CME_EO_resonant}
     \dot{b}_\nu = \frac{iJ_s}{2}\big(e^{i\theta}b_{\nu+s}e^{-i\frac{D_2s}{2}(2\nu+s)t} + e^{-i\theta}b_{\nu-s}e^{i\frac{D_2s}{2}(2\nu-s)t} \big).
\end{equation}

Now we aim to find the corresponding equation of the cavity field. In optical cavity the field envelop may be presented as Fourier series~\cite{herr2014temporal}
\begin{equation}
    A(\phi,t) = \sum_{\mu}a_\mu e^{i\mu \phi} = \sum_{\mu}b_{\mu}e^{i(\mu \phi - \omega_\mu t)}.
\end{equation}
Taking the time derivative, one obtains 
\begin{equation}
    \dot{A} = \sum_{\mu}(\dot{b}_\mu - i\omega_\mu b_\mu)e^{i(\mu \phi - \omega_\mu t)}.
\end{equation}

Let us consider only the first term. Substituting Eq.~(\ref{eq:CME_EO_resonant}) yields 
\begin{align}
   &\sum_{\mu} \dot{b}_\mu e^{i(\mu \phi - \omega_\mu t )} = i\sum_{\mu}e^{i(\mu \phi - \omega_\mu t )}\times\nonumber \\
   &\times J_{s}\big( a_{\mu-s}e^{i\frac{D_2s}{2}(2\mu-s)t +i\theta}+a_{\mu+s}e^{-i\frac{D_2s}{2}(2\mu+s)t - i\theta} \big).
\end{align}
One may readily rearrange the exponentials relations 
\begin{align*}
    \omega_{\mu-s} &= \omega_\mu - sD_1 + \frac{D_2}{2}(s^2 - 2\mu s)\\
    \omega_{\mu+s} &= \omega_\mu + sD_1 + \frac{D_2}{2}(s^2 + 2\mu s),
\end{align*}
and the summation yields that modulation creates a potential for the electric field 
\begin{equation}
    J_s\cos{(\phi s - sD_1t + \theta)}.
\end{equation}
Therefore, in the frame $\varphi$ rotating with speed $D_1$ such that $\varphi = \phi - D_1 t$, electric field obeys the following equation
\begin{equation}
    \dot{A} = i J_s\cos(s\varphi + \theta)A.
\end{equation}
This result might be combined with Lugiato-Lefever formalism for Kerr combs in optical cavities~\cite{Chembo2013Spatiotemporal}, and hence the equation which governs electric field in presence of external pump with frequency $\omega_p = \omega_0 + \delta\omega$ is
\begin{widetext}
\begin{equation}\label{eq:GPE}
    \frac{\partial A}{\partial t} = -\Big(\frac{\kappa}{2} + i\delta \omega\Big)A +\frac{i D_2}{2}\frac{\partial^2 A}{\partial \varphi^2}+ 2i J_s \cos\big(s \varphi + \theta\big)A + ig_0 \abs{A}^2 A + \sqrt{\kappa_{\mathrm{ex}}}s_{\mathrm{in}}.
\end{equation}
\end{widetext}

\section{Stable dispersionless limit}\label{sec:ap2}
Since the cubic equation~(\ref{eq:NL_NoDisp}) is written for real value $|\Psi|^2$, the solution has to be real as well. However, it is well known that a cubic equation always possesses three roots, and they are characterized through its discriminant $\Delta$. In our case, the discriminant has the following form 
\begin{align}\label{eq:discr}
    \Delta &= -27f^4 - 4(1 + \xi_0^2)^2 + 4f^2\xi_0(9 + \xi_0^2) + 4J \cos(s\varphi)\times\nonumber \\
    &\times\Bigg[-3f^2(3 + \xi_0^2) + 4\xi_0(1+\xi_0^2) - J \cos(s\varphi)\times\nonumber \\
    &\times\Big( 2 - 3f^2\xi_0 + 6\xi_0^2 + J\cos(s\varphi)\big(f^2 - 4\xi_0^2 + J\cos(s\varphi)  \big) \Big)   \Bigg].
\end{align}
Solving the equation $\Delta = 0$, we find values $f^2$ and $\zeta_0$ which determine the bi-stable zone. 

\section{Lagrangian pertubative approach}\label{sec:ap3}
In order to calculate the maximum detuning for DKS in a modulated cavity one needs to use the ansatz of a stationary soliton $\Psi = B e^{i\xi_0}\mathrm{sech}(B \Theta)$ in the Lagrangian density~(\ref{eq:Lagr_dens}). Integrating it over $\Theta$ on the interval $(-\infty,+\infty)$ (under the assumption $D_2/\kappa \ll 1$), one gets the Lagrangian in the form  
\begin{equation}
    L = -2B \frac{\partial \xi_0}{\partial \tau} + \frac{1}{3}B^3 -2B\xi_0 + \frac{J\alpha\pi }{\mathrm{sinh}\big(\frac{\alpha \pi}{2 B} \big)}. 
\end{equation}
The right hand side of Eq.~(\ref{eq:lagr_eq}) is not affected by the presence of the potential and coincides with works~\cite{grelu2015nonlinear,Wabnitz1993OpticalMem}. 
\bibliographystyle{apsrev4-1}

\bibliography{references}
\end{document}